\documentstyle{elsart}

\newcommand{ \bb }{ $2\nu\beta\beta \ $ }
\newcommand{ \bbm }{ $2\nu\beta^-\beta^- \ $ }

\begin{document}
\begin{frontmatter}
\title{THE STRENGTH OF THE ANALOG AND GAMOW-TELLER GIANT RESONANCES
AND HINDRANCE OF THE \bb - DECAY RATE}

\author[Moscow]{O.A.~Rumyantsev} and
\author[Moscow,Groningen]{M.H.~Urin \thanksref{e-mail}}
\address[Moscow]{Department of Theoretical Nuclear Physics,
Moscow Engineering Physics Institute, Moscow, 115409, Russia}
\address[Groningen]{Kernfysisch Versneller Institute, 9747AA Groningen,
The Netherlands}
\thanks[e-mail]{e-mail: urin@theor.mephi.msk.su, urin@kvi.nl}
\begin{abstract}
An approach for describing the hindrance of the nuclear \bb-decay 
amplitude is proposed. The approach is based on a new formula
obtained by a model-independent transformation of the initial
expression for the amplitude. 
This formula takes explicitly into account the hindrance of the decay-amplitude
 due to the presence of the collective Gamow-Teller state. 
Calculations are performed within the simplest version of the
approach.
Calculated and experimental \bb halve-lives are compared for
a wide range of nuclei.\\
\vspace{10pt}
\hspace{0. in}PACS numbers: 23.40-s ; 23.40.Hc ; 21.10.Tg\\
\hspace{0. in}Keywords: Double beta decay ; Collective states ; Half-live\\
\end{abstract}
\end{frontmatter}
\pagebreak

\section{Introduction}
A rough analysis of a great body of \bb-decay data
shows that the nuclear decay-amplitude is hindered as compared to 
the one evaluated within the independent-quasiparticle approximation.
(see e.g. refs \cite{Mor94,Amo94}). 
The hindrance is caused by the presence of the isobaric analog
and Gamow-Teller nuclear collective states (IAS and GTS)
at a rather large excitation energy. These states
exhaust a major fraction of, respectively, the Fermi and Gamow-Teller strength.
To describe the hindrance, different versions of the
quasiparticle random phase approximation (QRPA) are used
in  calculations of the nuclear \bb-decay amplitude. It has been
found (see e.g. refs.\cite{Vog88}-\cite{Civ95}) that the
calculated amplitudes are very sensitive to the particle-particle
interaction strength. For this reason the predictive power of the
current QRPA  versions is poor.
A similar instability of the calculated $0\nu\beta\beta$ -
decay amplitude has  also been found \cite{Hir95,Pan96}.\\
To bypass this difficulty, which probably is caused by an inconsistency
in the current versions of QRPA, we propose a rather different
 approach to describe the hindrance of the nuclear
\bb - decay amplitude.
The method is based on a new formula obtained by a 
model-independent transformation of the initial expression
for the amplitude. This formula takes explicitly into account
the decay-amplitude hindrance due to the presence of the collective 
GTS. The approach
is similar to the one widely used in the  description of
Fermi-type excitations in intermediate and heavy mass nuclei.
These excitations can be described with the explicit 
use of the
experimental fact, that the IAS exhausts  most (more than 95\%)
of the Fermi sum rule $N-Z$. For these nuclei the mean nuclear
Coulomb field is mainly responsible for the redistribution  
of the Fermi-strength
 from the IAS to its satellites.
If the variable part of this field were neglected, the IAS
would almost exhaust 100\% of the Fermi strength and the
Fermi \bb-amplitude would vanish for transitions to the ground and 
low-excited states of the product nucleus.
This approach, which allows one to evaluate the small Fermi-strength
of the IAS satellites populated virtually in the \bb-decay
process, has been originally proposed by Lane \cite{Lan61} and
is based on the  approximate conservation of isospin symmetry 
in nuclei.\\
A similar situation occurs in the description of Gamow-Teller
excitations. It is experimentally established that in intermediate
and heavy mass nuclei the GTS exhausts about 70\% - 80\% of the sum rule
$3e_q^2 ( N-Z ) $, where $e_q \simeq 0.8$ is ''an effective charge''
describing, in particular, the re-normalization of the axial coupling constant
for the weak interaction in nuclei. 
The GTS collectivity, directly related to  this experimental fact,
is explicitly used in the presented work to describe the hindrance
of the nuclear \bb-decay amplitude. Early attempts along this line have
been undertaken in refs. \cite{Ber90}-\cite{Rum96} within a perturbation
theory. 
As a first step in applying  the proposed approach, the decay 
amplitudes are calculated by the new formula using the BCS and the 
pair-vibration models to describe the corresponding states of
isobaric nuclei. Calculations performed for a wide range of nuclei
show clearly the hindrance of the decay amplitude due to the presence of
the collective GTS. 
The $2\nu\beta^-\beta^-$ and $2\nu\beta^+\beta^+$ (including
the electron capture) half-lives are
calculated using  model parameters
 taken from independent data. Results of calculations are compared
with available experimental data.

\section{The basic formula for the nuclear \bb - decay amplitude}
To make the following derivation of the new formula for the \bb-decay 
amplitude more transparent, we will use the same approach for
the Fermi and the Gamow-Teller transitions.
Let $G^{(\pm)}=\sum_a g_a^{(\pm)}$ be the operators of allowed
Fermi ($g^{(\pm)}=\tau^{(\pm)}$) and GT ($g^{(\pm)}=\vec\sigma
\tau^{(\pm)}$) $\beta$ - transitions. First, we consider
the {\bbm}-decay amplitude. Let $\vert i \rangle$ and $\vert f
\rangle$ be the ground state wave function, respectively,
of a double-even parent nucleus ($Z,N+2$) and of a product
nucleus ($Z+2,N$), $E_i$ and $E_f$ are the corresponding energies.
The expression for the nuclear \bb - decay amplitude has the
form (see e.g. refs. \cite{Doi88,Doi92}):

\begin{equation}
\label{Mbb}
M_G = \sum_S {\langle f \vert G^{(-)}\vert S\rangle\langle S\vert
G^{(-)} \vert i \rangle}{\omega_S^{-1}}\ ,
\end{equation}

\noindent where the sum is taken over intermediate states of an
isobaric nucleus $(N+1,Z+1)$ and $\omega_S = E_S-(E_i + E_f)/2$
is the excitation energy of an intermediate state. The
$2\nu\beta^+ \beta^+$ - decay amplitude is equal to conjugate
value $M_G^* $.

We transform eq.(\ref{Mbb}) in the following way.
Let $\vert G\rangle$ be one of the intermediate states with the energy
$E_G$ and $V_G^{(-)}$ be the following operator:

\begin{equation}
\label{VG}
V_G^{(-)}=-i\hbar\dot G^{(-)} - \Delta_G G^{(-)},\ \
\Delta_G = E_G - E_i\ ,
\end{equation}

\noindent where a dot over the operator implies the time derivative.
Matrix elements of this operator can readily be constructed:

\begin{eqnarray}
\label{delta}
\nonumber -({\omega_G-\omega_S})\langle S\vert G^{(-)}\vert i\rangle = {\langle S\vert V_G^{(-)}
\vert i\rangle}\ ,\\
-({\omega_G+\omega_S})\langle f\vert G^{(-)}\vert S\rangle={\langle f\vert V_G^{(-)}
\vert S\rangle}\ ,
\end{eqnarray}

\noindent 
giving
$\langle G\vert V_G^{(-)}\vert i\rangle = 0$. Using these
matrix elements we transform amplitude (\ref{Mbb}) to the
expression:

\begin{equation}
M_G = \sum_{S\ne G} \frac{\langle f \vert V_G^{(-)} \vert S\rangle\langle
S\vert V_G^{(-)} \vert i \rangle}{\omega_S(\omega_G^2-\omega_S^2)} -
\frac{\langle f \vert V_G^{(-)} \vert G\rangle\langle G\vert G^{(-)}
\vert i\rangle}{2\omega_G^2}\ .
\end{equation}

\noindent The last term in this expression can be transformed
with the help of eqs.(\ref{VG}),(\ref{delta}) 
and the equalities

\begin{equation}
\label{eq}
0=\langle f\vert [V_G^{(-)},G^{(-)}]\vert i\rangle=
\sum_S\Bigl(
\langle f\vert V_G^{(-)}\vert S\rangle
\langle S\vert G^{(-)}\vert i\rangle -
\langle f\vert G^{(-)}\vert S\rangle
\langle S\vert V_G^{(-)}\vert i\rangle\Bigr).
\end{equation}

\noindent The commutator vanishes because of the fact 
that $(\tau^{(-)})^2=0$.
As a result of the transformations 
performed with the help of  eqs.(\ref{VG})-(\ref{eq})  we get the final expression for
the nuclear \bb - decay amplitude:

\begin{equation}
\label{MG}
M_G =\omega_G^{-2} \sum_S {\langle f \vert V_G^{(-)} \vert S
\rangle \langle S \vert V_G^{(-)} \vert i \rangle}{\omega_S^{-1}}\ .
\end{equation}

\noindent Summation in this equation formally includes the  state $\vert G\rangle$
as well, because $\langle G\vert V_G^{(-)}\vert i\rangle = 0$.

Eqs. (\ref{Mbb}) and (\ref{MG}) are equivalent. However,
the \bb-decay hindrance due to the presence of the collective states
(IAS and GTS) can be explicitly taken into account in eq.(\ref{MG}) provided that the state
$\vert G\rangle$ is, respectively, taken as one of these
states. Indeed,  if the above states each exhaust 100\% of the
corresponding particle-hole strength, i.e. $\vert G\rangle \sim
G^{(-)}\vert i\rangle$, then matrix elements $\langle S\vert V_G^{(-)}
\vert i \rangle$ and, therefore, the amplitudes $M_G$ are equal to zero.
It is now clear, in a model-independent way, that if the collective state $\vert G \rangle$ 
exhausts most of the sum-rule strength, the \bb-decay will be strongly
suppressed.
 
For the above reasons the use of eq.(\ref{MG}) allows one to bypass
the difficulty connected with the use
of  current versions of QRPA which are based on  eq.(\ref{Mbb}). 
To show it we first
 consider nuclei without nucleon pairing using the phenomenological nuclear
mean field $U$ and the isovector part of the Landau-Migdal particle-hole
interaction $F$. The mean field contains isoscalar, isovector,
spin-orbit and Coulomb parts:

\begin{equation}
\label{Ushell}
U=U_0(r)+U_\tau(r)\tau^{(3)}+U_{\sigma l}(r)(\vec\sigma\vec l)+
U_C(r)(1-\tau^{(3)})/2\ .
\end{equation}

\noindent The interaction

\begin{equation}
\label{LM}
F=(F_\tau+F_{\sigma\tau}\vec\sigma_1\vec\sigma_2)
\vec\tau_1\vec\tau_2\delta(\vec r_1-\vec r_2)
\end{equation}

\noindent contains two phenomenological parameters $F_g 
(g=\tau,\sigma\tau)$ which are widely used within in the theory of finite Fermi-systems
\cite{Mig83}. Let $\rho(r)=(\rho^{(+)}(r)+\tau^{(3)} \rho^{(-)}(r))/2$ be nucleon
density: $\rho^{(\pm)}(r)=\rho^n(r)\pm\rho^p(r)$, $\rho^n$ and $\rho^p$
are the neutron and proton densities, respectively. The following
conditions

\begin{equation}
U_\tau(r)=F_\tau\rho^{(-)};\ \ U_C(r)=e^2\int\rho^p(r_1)\vert\vec r-\vec r_1
\vert^{-1}d\vec r_1
\end{equation}

\noindent allow one to calculate two terms in eq.(\ref{Ushell}) in
a self-consistent way (the first condition is discussed in  e.g. ref.\cite{Bir74}).

Let us turn to the operator $V_G^{(-)}=\sum_a V_g^{(-)}(a)$ (\ref{VG}). It can
be shown using the coordinate representation of the RPA equations that:

\begin{eqnarray}
\label{RPA}
\nonumber 
-i\hbar\dot g^{(-)}=[h,g^{(-)}]-2F_g[\rho,g^{(-)}]\ ,\\
V_g^{(-)}=[h,g^{(-)}]-2F_g[\rho,g^{(-)}]-\Delta_G g^{(-)}.
\end{eqnarray}

\noindent Here, $h=t+U$ is the single-particle Hamiltonian.
Assuming  $\vert G\rangle\sim G^{(-)}\vert i\rangle$,
we find an approximate expression for the energy $\Delta_G$
using the equality $\langle G\vert V_G^{(-)}\vert i\rangle = 0$
and eq.(\ref{VG}):

\begin{equation}
\label{deltaE}
\Delta_G \cong -i\hbar\frac{\langle i\vert [G^{(+)},\dot G^{(-)}]
\vert i\rangle}{\langle i\vert
[G^{(+)},G^{(-)}]\vert i\rangle}\ .
\end{equation}

\noindent This expression can also be obtained by means of the energy
weighted sum rule where $\Delta_G$  equals to the mean energy of Fermi or GT 
excitations, respectively. This formula has acceptable accuracy because
the IAS and GTS exhaust most of the corresponding particle-hole strength.

For the case of Fermi transitions ($g^{(\pm)}=\tau^{(\pm)}$) we
obtain from eqs.(\ref{Ushell})-(\ref{deltaE}):

\begin{equation}
\label{DC}
V_\tau^{(-)}=(U_C(r)-\Delta_C)\tau^{(-)};\ \
\Delta_C=(N-Z+2)^{-1}\int U_C(r)\rho^{(-)}(r)\ d\vec r\ .
\end{equation}

\noindent These well-known equations are a result of
the approximate conservation of isospin symmetry in medium and
heavy nuclei, where the mean Coulomb field is the main source
of violation of this symmetry. Similar calculations 
for the case of Gamow-Teller transitions ($g^{(\pm)}=\vec\sigma
\tau^{(\pm)}$) lead to the following expression:

\begin{eqnarray}
\label{Vg}
\nonumber 
V_{\sigma\tau}^{(-)}=(U_C(r)-\Delta_C)\vec\sigma\tau^{(-)}+
(U_{\sigma l}(r)[\vec\sigma\vec l,\vec\sigma]-\Delta_{\sigma l}
\vec\sigma)\tau^{(-)}+\\
(2 \delta F\rho^{(-)}(r)-\Delta_{\delta F})\vec\sigma\tau^{(-)},
\end{eqnarray}

\noindent where $\Delta_C$ is determined by eqs.(\ref{DC}),
$\delta F=F_\tau-F_{\sigma\tau}$,

\begin{eqnarray}
\label{DS}
\nonumber 
\Delta_{\sigma l}=-\frac{2}{3}(N-Z+2)^{-1}
\langle i\vert\sum_a U_{\sigma l}(r_a)\vec\sigma_a\vec
l_a\vert i\rangle,\\
\Delta_{\delta F}=2(N-Z+2)^{-1}\delta F
\int(\rho^{(-)}(r))^2\ d\vec r
\end{eqnarray}

\noindent and

\begin{equation}
\label{DG}
\Delta_{\sigma\tau}=\Delta_C+\Delta_{\sigma l}+\Delta_{\delta F}\ .
\end{equation}

\noindent The three terms in eq.(\ref{Vg}) arise as a result of 
spin-isospin symmetry violation (this symmetry may be called
 a "projection on the charge-exchange channel" of the spin-isospin
SU(4) - symmetry \cite{Gap92}). However, due to the smooth radial
dependence of both the mean Coulomb field and the excess neutron
density the spin-orbit part of the nuclear mean field is the
main source of the redistribution of 
the Gamow-Teller strength from the
GTS to its satellites.
In other words, only the second term in eq.(\ref{VG}) needs to be taken
into consideration in the analysis of the Gamow-Teller
\bb - decay amplitude by means of eq.(\ref{MG}). 
The other terms are only important for the calculation of the 
energy of the GTS (\ref{DG}).

The equations equivalent to eqs.~(\ref{RPA}) have not yet been formulated  for
nuclei with pairing. Nevertheless, one can state that
the operator $V_\tau^{(-)}$ (\ref{DC})
is not changed 
since  paring forces conserve isospin symmetry. In the case
of Gamow-Teller transitions the operator $V_{\sigma \tau}^{(-)}$ (\ref{Vg})
is expected to be somewhat modified due to that part of the
particle-particle interaction, which violates the ``projection''
of SU(4) - symmetry.
 The relative contribution of this interaction to
$V_{\sigma \tau}^{(-)}$ can be roughly estimated as the ratio
 of the energy gap to the mean single-particle
spin-orbit energy-splitting, which is rather small (no more than 
20\%). This contribution is omitted in the following
analysis.

In view of a higher degree of the isospin-symmetry conservation
as compared with the spin-isospin-symmetry conservation one can conclude
that $M_{GT}\gg M_F$. For this reason we consider in the following
only the Gamow-Teller \bb - decay amplitude.

\section{Evaluation of the decay amplitude within a simplest
version of the approach}
As a first step in evaluating  the amplitude $M_{GT}$ with
the help of the eq.(\ref{MG}) we use the simplest approximation
for describing the corresponding states in  isobaric nuclei.
 Namely, we use the BCS model for describing
subsystems with strong nucleon pairing and the pair-vibration
model for describing "magic $\pm$ two nucleons" subsystems.

Within this approximation the nucleon densities $\rho^{(\pm)}(r)$
are equal to

\begin{equation}
\rho^{(\pm)}(r) ={1\over 4\pi}\left(\sum_\nu R^2_\nu(r)(2j_\nu+1)n_\nu
\pm \sum_\pi R^2_\pi (r) (2j_\pi + 1) n_\pi\right ),
\end{equation}

\noindent where $R_\lambda (r)$ are the radial single-neutron
($\lambda = \nu$) and single-proton ($\lambda = \pi$) wave functions;
$\lambda = n_r, l, j$ is the set of the single-particle quantum
numbers, and $n_\lambda$ are occupation numbers satisfying the equations:

\begin{equation}
\label{NN}
\sum_\nu (2j_\nu + 1)n_\nu = N + 2,\ \ \ \sum_\pi (2j_\pi + 1)n_\pi =Z.
\end{equation}

\noindent For nuclei with strong pairing in any nucleon subsystem
the occupation numbers are $n_{\lambda} = v^2_\lambda = 1 -
u^2_\lambda$, where $v_\lambda$ and $u_\lambda$ are the
Bogoliubov-transformation coefficients. In the case that a nucleon
subsystem is the ''magic $\pm$ two nucleons'' subsystem the occupation factors
are: $n_\lambda = n^m_\lambda + c^2_\lambda(1-2 n^m_\lambda)$,
where $n^m_\lambda$ are the occupations numbers for the magic
subsystem; $c_\lambda$ are the coefficients, which  determine the
pair-vibration wave function and satisfy the normalization
condition: $\sum_\lambda c^2_\lambda (1-2n^m_\lambda )(2j_\lambda
+ 1) = \pm 2$.

The terms $\Delta_C,\ \Delta_{\sigma l}$ and $\Delta_{\delta F}$
determining the GTS energy (\ref{DG}) can also be calculated easily
according to eqs.(\ref{DC}),(\ref{DS}). In particular, we have

\begin{equation}
\Delta_{\sigma l}=-\frac{8}{3}(N-Z+2)^{-1}\sum_{\lambda=\pi,\nu}
n_\lambda\langle\lambda\vert U_{\sigma l}(r)\vert\lambda\rangle
(j_\lambda-l_\lambda)l_\lambda(l_\lambda+1),
\end{equation}

\noindent where $\langle\pi\vert U_{\sigma l}(r)\vert\nu\rangle=
\int R_\pi(r)R_\nu(r) U_{\sigma l}(r)r^2\ dr$ and $n_\lambda$ are
occupation numbers satisfying eqs.(\ref{NN}).

Calculation of the amplitude $M_{GT}$ according to eq.(\ref{MG})
within the framework of the BCS model results in the expression:

\begin{eqnarray}
\label{MGT}
\nonumber
M_{GT} =e^2_q \omega_{GTS}^{-2} {\displaystyle \sum_{\pi,\nu}}
\Biggl( (2l_\pi+1)(j_\pi-j_\nu) \langle \pi \vert U_{SO}(r) \vert \nu
\rangle - \Delta_{SO} \langle \pi \vert \nu \rangle \Biggr)^2 \\
\times \langle \pi
\Vert \sigma \Vert \nu \rangle^2 u_\pi v_\pi u_\nu v_\nu
\omega_{\pi \nu}^{-1}\ .
\end{eqnarray}

\noindent Here, $\langle\pi\Vert\sigma\Vert\nu\rangle$ is the reduced
matrix element; $\omega_{\pi\nu}={\cal E}_\pi+{\cal E}_\nu$ is the
excitation energy of the two-quasiparticle state, $ {\cal E}_\lambda =
\sqrt{(\epsilon_\lambda - \mu )^2 + \Delta^2}$ is the
single-quasiparticle energy for the subsystem with strong nucleon
pairing, $\epsilon_\lambda$ is the energy of single-particle level,
and $\Delta$ is the energy gap. If the nucleon subsystem is magic in final
(initial) state and ''magic $\pm$ two nucleon'' in initial (final)
state, 
we use $c_\lambda$ instead of $u_\lambda v_\lambda$
 in eq. (\ref{MGT})
and ${\cal E}_\lambda =
\vert \Delta + \epsilon_\lambda - \epsilon_1 \vert$ for 
particle pair-vibrations or ${\cal E}_\lambda=\vert \Delta
+ \epsilon_0 - \epsilon_\lambda \vert$ for hole pair-vibrations,
where $\epsilon_1$ ($\epsilon_0$) is the energy of the first empty
(last filled) single-particle level in the magic subsystem. Here, $2\Delta$
is the pair-vibration state energy, which coincides with the pairing energy
$P$ for the subsystem ''magic $\pm$ two nucleons''. The hindrance factor
$h_{GT}$ can be estimated as follows: $h_{GT} = M_{GT}/M_{GT}^0$, where
amplitude

\begin{equation}
\label{MGT0}
M_{GT}^0 = e^2_q \sum_{\pi,\nu} \langle \pi \vert \nu \rangle^2 \langle
\pi \Vert \sigma \Vert \nu \rangle^2 u_\pi v_\pi u_\nu v_\nu
\omega_{\pi\nu}^{-1}\ .
\end{equation}

\noindent is evaluated according to the initial equation (\ref{Mbb})
with the use of the BCS (or BCS and pair-vibration) model without
consideration of the hindrance caused by the presense of the
collective state.

\section{Results of the calculations and summary}
 The parameterization  of the
isoscalar part of the phenomenological nuclear mean field including the spin-orbit
term is  given e.g. in ref. \cite{Che67}. The strength
$F_\tau= 300\ MeV\ fm^3$ in eq. (\ref{LM}) is chosen to describe
the experimental neutron and proton binding-energy difference
for nuclei with rather large neutron excess $^{48}Ca,^{68}Ni, ^{132}Sn,
^{208}Pb$ \cite{Wap85}, for which the difference is mainly determined
by the symmetry potential and mean Coulomb field. The strength
$F_{\sigma \tau}= 255\ MeV\ fm^3$ in eq.(\ref{LM}) is chosen following the suggestion 
of  ref.
\cite{Mig83}.
The parameters $\mu_{n,p}$ and $\Delta_{n,p}$, 
obeying the constraints of eqs. (\ref{NN}), were
found for each subsystem with strong nucleon pairing 
to describe the experimental pairing
energies $P = 2 \min_\lambda {\cal E}_\lambda$ taken according
to ref. \cite{Wap85}. The pair-vibration state energies
$2\Delta=P$ are calculated using experimental pairing energies
\cite{Wap85}. Coefficients $c_\lambda$ determining the
pair-vibration state wave function for a nucleon subsystem
''magic + or -- two nucleons'', are calculated using
 $c_\lambda\sim(1-2n^m_\lambda)/(\epsilon_\lambda -
\epsilon_1 + 2\Delta ) $ or $c_\lambda\sim (1-2n^m_\lambda)/
(\epsilon_0 -\epsilon_\lambda + 2\Delta)$ with taking into
account the normalization conditions given above \cite{Sap85}.
For all considered nuclei the single-particle basis including
all bound states as well as the quasi-bound states up to
$\sim 5 MeV$ is used. If the nucleon subsystem in a parent nucleus
is ''magic $\pm$ two nucleons'', and in the product-nucleus is
''magic $\pm$ four nucleons'', the calculations were performed
within the framework of the BCS model with the use of $u_\lambda,
v_\lambda$ factors calculated for a ''magic $\pm$ four nucleons''
subsystem.

Results of the calculations and corresponding experimental data are given
in Table 1. The $M_{GT}$ amplitudes calculated using eq.(\ref{MGT})
are given in column 5.
The $M^0_{GT}$ amplitudes were calculated from eq. (\ref{MGT0})
with the use of the same model parameters and the same number
of basis states. The calculated hindrance factors $h_{GT}$
are given in column 6. Half-lives $T_{1/2}^{calc}$ calculated
by formula $(T_{1/2})^{-1}=G_{2\nu}
\vert M_{GT}\vert^2$ are given in column 8. The lepton factors
$G_{2\nu} $ taken from refs. \cite{Amo94,Doi88,Doi92} are also
given (column 3). We calculated also the $2\nu\beta^+\beta^+$
half-lives (including the electron capture) for those nuclei,
for which  relevant experimental data is expected to be forthcoming.

In this work the initial formula
for the nuclear \bb - decay amplitude is transformed in a
model-independent way to an expression, which
takes explicitly into account the decay-rate hindrance due to the
presence of the collective GTS. To apply the transformed formula
in its simplest version we use: (i) the fact that the GTS
exhausts the most of the Gamow-Teller strength; (ii)
the BCS and the pair-vibration models for describing the corresponding
states of  isobaric nuclei. 
The formula for the amplitude $M_{GT}$, obtained with the use of 
the above approximations ,
is the sum of the positive-sign terms and, therefore, is stable
to reasonable variations of model parameters. Calculations are performed
for a wide range of nuclei with the use of the parameters
found from independent data. Except for $Te$ isotopes the
calculated amplitudes $M_{GT}$ within the factor 2--3 are
in agreement with the corresponding amplitudes $M_{GT}^{exp}$
deduced from experimental data.

The present approach for evaluating the nuclear \bb-decay amplitude
is  incorporating in a model-independent way the mechanism for 
the rather strong suppression of the decay-rate and is thus to be 
preferred over conventional approaches. The use of the present approach
in QRPA requires further investigation.

\section{Acknowledgements}
The authors are grateful to L.Dieperink, S.A.Fayans,  Yu.V.Gaponov
and O.~Scholten for discussions. One of the authors
(M.H.U.) acknowledges generous financial support from 
the ``Nederlandse organisatie voor wetenschappelijk onderzoek''
NWO during his stay at the KVI.

\newpage
\textwidth=19cm
\oddsidemargin=-1.5cm
Table 1. Calculated $M_{GT}$ and $T_{1/2}$ values in comparison with the
corresponding experimental data. $^{150}$Nd is considered as a 
spherical nucleus. The lepton factors $G_{2\nu}$ and calculated hindrance factors
are also given. $m_e$ is the electron mass. 
\medskip

\begin{tabular}{|c|c|cc|c|c|c|cc|c|}
\hline
parent    & type of    & $G_{2\nu}$   & & $M_{GT}^{exp.}$ & $M_{GT}$  & $h_{GT}$
& $T_{1/2}^{exp.}$     &              & $T_{1/2}^{calc.}$    \\
nucleus   & decay      & $years^{-1}\ m_e^2$ &  & $m_e^{-1}$ & $m_e^{-1}$  &
& $years$              &              & $years$                \\
\hline
$^{76}Ge $  & $\beta^-\beta^-$ & $1.317\times 10^{-19}$ & \cite{Doi88} & 0.0919           & 0.0387            & 0.131   & $0.9 \times 10^{21}$ & \cite{Vas90} & $5.0 \times 10^{21}$ \\
            &                  &                        &              & 0.0737           &                   &         & $1.43\times 10^{21}$ & \cite{Bal92} &                      \\
$^{78}Kr $  & $ec\ ec$         & $1.957\times 10^{-21}$ & \cite{Doi92} &                  & 0.0285            & 0.090   &                      &              & $6.2 \times 10^{23}$ \\
            & $\beta^+ec$      & $1.174\times 10^{-21}$ & \cite{Doi92} &                  &                   &         &                      &              & $1.0 \times 10^{24}$ \\
$^{82}Se $  & $\beta^-\beta^-$ & $4.393\times 10^{-18}$ & \cite{Doi88} & 0.0459           & 0.0295            & 0.093   & $1.08\times 10^{20}$ & \cite{Ell92} & $2.6 \times 10^{20}$ \\
$^{96}Zr $  & $\beta^-\beta^-$ & $1.953\times 10^{-17}$ &              & 0.0362           & 0.0678            & 0.324   & $3.9 \times 10^{19}$ &              & $1.1 \times 10^{19}$ \\
$^{96}Ru $  & $ec\ ec$         & $6.936\times 10^{-21}$ & \cite{Doi92} &                  & 0.1005            & 0.338   &                      &              & $1.4 \times 10^{22}$ \\
            & $\beta^+ec$      & $1.148\times 10^{-21}$ & \cite{Doi92} &                  &                   &         & $>6.7\times 10^{16}$ & \cite{Nor84} & $8.6 \times 10^{22}$ \\
$^{100}Mo$  & $\beta^-\beta^-$ & $9.553\times 10^{-18}$ & \cite{Doi88} & 0.0954           & 0.1606            & 0.329   & $1.15\times 10^{19}$ & \cite{Eji91} & $4.1 \times 10^{18}$ \\
$^{106}Cd$  & $ec\ ec$         & $1.573\times 10^{-20}$ & \cite{Doi92} &                  & 0.1947            & 0.319   &                      &              & $1.7 \times 10^{21}$ \\
            & $\beta^+ec$      & $1.970\times 10^{-21}$ & \cite{Doi92} &                  &                   &         & $>6.6\times 10^{18}$ & \cite{Barab} & $1.3 \times 10^{22}$ \\
$^{116}Cd$  & $\beta^-\beta^-$ & $8.000\times 10^{-18}$ & \cite{Amo94} & 0.0754           & 0.0788            & 0.258   & $2.25\times 10^{19}$ & \cite{Amo94} & $1.2 \times 10^{19}$ \\
$^{124}Xe$  & $ec\ ec$         & $5.101\times 10^{-20}$ & \cite{Doi92} &                  & 0.0528            & 0.085   &                      &              & $7.0 \times 10^{21}$ \\
            & $\beta^+ec$      & $4.353\times 10^{-21}$ & \cite{Doi92} &                  &                   &         &                      &              & $8.2 \times 10^{22}$ \\
$^{128}Te$  & $\beta^-\beta^-$ & $8.624\times 10^{-22}$ & \cite{Doi88} & 0.0123           & 0.0529            & 0.090   & $7.7 \times 10^{24}$ & \cite{Ber93} & $4.1 \times 10^{23}$ \\
$^{130}Te$  & $\beta^-\beta^-$ & $4.849\times 10^{-18}$ & \cite{Doi88} & 0.0087           & 0.0468            & 0.085   & $2.7 \times 10^{21}$ & \cite{Ber93} & $9.4 \times 10^{19}$ \\
$^{130}Ba$  & $ec\ ec$         & $4.134\times 10^{-20}$ & \cite{Doi92} &                  & 0.0568            & 0.082   & $>4  \times 10^{21}$ & \cite{Bar95} & $7.5 \times 10^{21}$ \\
            & $\beta^+ec$      & $1.387\times 10^{-21}$ & \cite{Doi92} &                  &                   &         & $>4  \times 10^{21}$ & \cite{Bar95} & $2.2 \times 10^{23}$ \\
$^{136}Xe$  & $\beta^-\beta^-$ & $4.870\times 10^{-18}$ & \cite{Doi88} & 0.0299           & 0.0341            & 0.088   & $>2.3\times 10^{20}$ & \cite{Vui93} & $1.1 \times 10^{20}$ \\
$^{136}Ce$  & $ec\ ec$         & $3.988\times 10^{-20}$ & \cite{Doi92} &                  & 0.0512            & 0.081   &                      &              & $9.6 \times 10^{21}$ \\
            & $\beta^+ec$      & $6.399\times 10^{-22}$ & \cite{Doi92} &                  &                   &         &                      &              & $6.0 \times 10^{23}$ \\
$^{150}Nd$  & $\beta^-\beta^-$ & $1.200\times 10^{-16}$ & \cite{Doi88} & 0.0221           & 0.0642            & 0.182   & $1.7 \times 10^{19}$ & \cite{Art93} & $2.0 \times 10^{18}$ \\
            &                  &                        &              & 0.0304           &                   &         & $9   \times 10^{18}$ & \cite{Moe92} &                      \\
\hline
\end{tabular}
The table should be put in the end of Sect. 4
\end{document}